\documentclass[twocolumn,prb,aps]{revtex4}

\usepackage{graphicx}
\usepackage{dcolumn}
\usepackage{bm}
\usepackage{epsfig}
\usepackage{endnotes}

\begin{document}


\title{Microwave Dielectric Loss at Single Photon Energies and milliKelvin Temperatures}
\author{Aaron D. O'Connell}
\author{M. Ansmann}
\author{R. C. Bialczak}
\author{M. Hofheinz}
\author{N. Katz}
\author{Erik Lucero}
\author{C. McKenney}
\author{M. Neeley}
\author{H. Wang}
\author{E. M. Weig}
\author{A. N. Cleland}
\author{J. M. Martinis}
\affiliation{Department of Physics, University of California, Santa
Barbara, CA 93106}

\date{\today}

\begin{abstract}
The microwave performance of amorphous dielectric materials at very low temperatures and very low
excitation strengths displays significant excess loss. Here, we present the loss tangents of some
common amorphous and crystalline dielectrics, measured at low temperatures ($T <$ 100 mK) with near
single-photon excitation energies, $E/\hbar\omega_0 \sim 1$, using both coplanar waveguide (CPW)
and lumped $LC$ resonators. The loss can be understood using a two-level state (TLS) defect model.
A circuit analysis of the half-wavelength resonators we used is outlined, and the energy
dissipation of such a resonator on a multilayered dielectric substrate is considered theoretically.
\end{abstract}

\maketitle

Dielectric loss is a significant concern for superconducting quantum bits (qubits), as energy
relaxation within the dielectric is one of the primary sources of quantum
decoherence\cite{Martinis:2005}.  Superconducting qubits operate in the low-temperature,
low-voltage regime, where dielectric loss is typically not well characterized. While the dielectric
loss may be extremely small at higher excitation voltages and temperatures, it has been observed
that the loss tangent scales inversely with voltage ($\tan\delta\sim 1/V_{rms}$) and levels off at an
intrinsic value $\tan\delta_i$ that is often substantially greater than the loss at larger
voltages, as shown in Fig. 1. The lowest excitation voltages shown there correspond to of order 1
photon in a 6 GHz $LC$ resonator, with $C \sim 1$ pF.

\begin{figure}
\includegraphics[width=0.48\textwidth]{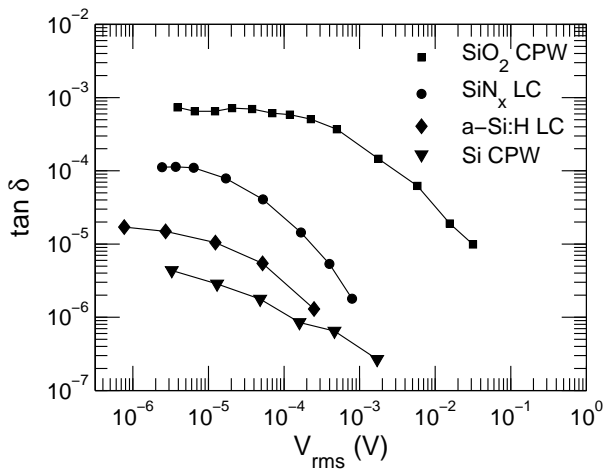} \caption{Loss tangent after adjusting
for the electrical loading. Data labeled $\textrm{SiO}_2$ and Si correspond to 300nm PECVD
$\textrm{SiO}_2$ on single-crystal Si, and 100 $\Omega$-cm single-crystal Si, respectively. All
resonators had Al electrodes. T $\leq$ 100 mK.} \label{fig:figure1}
\end{figure}

This behavior has been postulated to arise from coupling to a bath of TLS defects in the
dielectric, which absorb and disperse energy at low power but become saturated with increasing
voltage and temperature \cite{Shnirman:2005}.  TLS are found in most amorphous materials and arise
from an energy difference between defect bond configurations coupled by tunneling.  The bath of TLS
is assumed to have a constant distribution in energy and a log uniform distribution in transition
strength \cite{Hunklinger:1986}.  These defect states couple to the surrounding electric field
through the electric dipole moments that arise from differences in the charge distribution between
configurations \cite{Schickfus:1977}.

Although dielectric loss at higher powers and temperatures has been extensively reported, the literature
contains very little information on dielectric performance in the low-temperature, low-voltage limit.  Guided by prior measurements of
hydrogenated dielectrics with over-constrained lattices \cite{Pohl:2002}, we examined the microwave
loss of a range of dielectric materials compatible with qubit fabrication.  Here we report direct
measurements of the intrinsic loss tangents of these dielectric materials.

To perform these measurements, we fabricated both parallel $LC$ resonators, comprising a
superconducting inductive coil and a parallel-plate capacitor containing the dielectric in
question, and half-wavelength CPW resonators, where the single-layer superconducting metal
electrodes are patterned atop the dielectric. A CPW resonator is shown in Fig. 2(a). $LC$ resonators
afford more straightforward analysis of the loss tangent,  due to the parallel electric field
configuration between the plates, while CPW resonators are easier to fabricate, but require more
complicated analysis. Both types of resonators were coupled to measurement lines through on-chip
coupling capacitors $C_c$, as illustrated in Fig. 2(b). The resonators had resonance frequencies
near the 6 GHz operating frequencies of our qubits. The resonators' transmission
$S$-parameter, $S_{21}$, was measured as a function of voltage and temperature, using a vector
network analyzer. The loss tangents were extracted as described below. The results of these
measurements are compiled in Table I.

\begin{table}
  \caption{Intrinsic loss tangents, after accounting for external loss and CPW field-distribution analysis.
  Deposited films have typical thickness of a few hundred nanometers. Materials marked ``SC'' indicate single crystals.}
  \begin{ruledtabular}
  \begin{tabular}{l c c c}
  Dielectric & Metal & Resonator & $\tan \delta_i \times 10^6$ \vspace{3 pt} \\
  \hline \\
  100 $\Omega$-cm Si (SC) & Al & CPW & $<$ 5$-$12 \\ 
  Sapphire (SC) & Re & CPW & $<$ 6$-$10 \\ 
  Sapphire (SC) & Al & CPW & $<$ 9$-$21 \\ 
  $a$-Si:H & Al & $LC$ & 22$-$25\\ 
  $a$-Si:H & Al & CPW & 10$-$130 \\ 
  Interdigitated cap. & Al & $LC$ & 41$-$47\\ 
  \hspace{2 mm} on sapphire (SC) & & & \\
  SiN$_x$ & Re or Al & $LC$ or CPW & 100$-$200\\ 
  Thermal SiO$_2$ & Al & CPW & 300$-$330 \\ 
  Sputtered Si & Al & CPW & 500$-$600 \\ 
  AlN & Al & CPW & 1100$-$1800\\ 
  PECVD SiO$_2$ & Al & CPW & 2700$-$2900\\ 
  MgO & Al & CPW & 5000$-$8000 \\ 
  \end{tabular}
  \end{ruledtabular}
\end{table}

Near its half-wave resonance frequency, a CPW resonator can be represented by an equivalent $LC$
lumped circuit, shown in Fig. 2(b). The Norton equivalent circuit is shown in Fig. 2(c), where the voltage source has been transformed to a
current bias $V_1/(R_0 + Z_c)\simeq V_1/Z_c$, and the impedance $R_0 + Z_c$ can be written as $Z_c \parallel
|Z_c|^2/R_0$, where we have used $|Z_c| = 1/\omega C_c \gg R_0$ for typical coupling capacitances
$C_c$ on the order of a few fF.  This can now be viewed as a parallel $LCR$ circuit with effective capacitance
$C' = C + 2C_c$ and resistance $R' = R\parallel |Z_c|^2/2R_0$ (Fig. 2(d)). The response at
frequency $\omega$, near the resonance frequency $\omega_0 = 1/\sqrt{L C'}$, is given by
\begin{equation}
V = \frac{V_1}{Z_c}\frac{1}{1/R'+1/i\omega L+i\omega C'}.
\end{equation}
The output voltage $V_2$, as shown in Fig. 2(b), is given by $V_2 = VR_0/(R_0+Z_c)\simeq VR_0/Z_c$.
The normalized scattering matrix parameter is given by  $S_{21} = 2V_2/V_1$, where we have used
$|S_{21}|=1$ for the on-resonance transmittance of a lossless resonator.  Finally, taking $Q_m =
R'/\omega_0L$, $R_c = |Z_c|^2/2R_0$, and assuming $Q_m\gg1$,  we obtain
\begin{equation}
\label{eq:S21}
S_{21} \simeq -\frac{1}{1+R_c/R}\frac{1}{1+i2Q_m(\omega-\omega_0)/\omega_0}.
\end{equation}

This equation is used to fit our measured $S_{21}$ data, from which we can extract the total measured quality factor $Q_m = 1/\tan\delta$.  The quality factor is attributed to the parallel sum of two
independent loss mechanisms, $1/Q_m = 1/Q_0 + 1/Q_c$, where $1/Q_0$ is the internal dielectric loss,
and $1/Q_c$ the loss due to the measurement impedance $R_0$. We calculate $Q_c$ either from
the formula $1/Q_c = 2 R_0 Z_0 \omega_0^2 C_c^2$, where $Z_0$ is the resonator characteristic impedance, or
through the relation $Q_c = Q_m/|S_{21}|$ for over-coupled samples, when $Q_m$ saturates at high powers and $ |S_{21}|\simeq 1$. Finally, the limiting loss tangent is related to $Q_0$ at the lowest excitation voltage, $\tan\delta_0 = 1/Q_0$.

For an $LC$ resonator, this limiting loss tangent is a direct measurement of the low-power,
low-temperature intrinsic loss of the dielectric, $\tan \delta_i$.  This can be seen by noting that
the electric field in an $LC$ resonator is almost entirely confined to the space between the
capacitor plates.  Furthermore, the inductive loss is typically negligible at these temperatures\cite{Mazin:2002}. However, in a CPW resonator, the electric field samples a large volume of space around the CPW not filled by the
dielectric of interest, so the limiting loss tangent $\tan \delta_0$ is not identical to the
intrinsic loss tangent. For a CPW resonator fabricated on a multi-layer substrate, it is necessary
to know the fraction of the electrical energy stored in each dielectric, and the intrinsic loss
tangents for all but one of the constituent dielectrics, as well as the value of limiting loss
tangent for the composite structure.

\begin{figure} \centering
\includegraphics[width=0.45\textwidth]{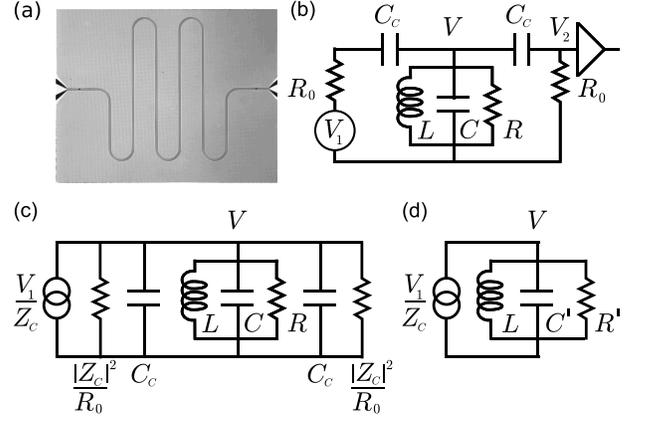} \caption{(a) Micrograph of a half-wavelength CPW
resonator.  (b) Circuit representation and measurement lines. (c) Norton equivalent circuit ($|Z_c|
\gg R_0$). (d) $LCR$ equivalent circuit.} \label{fig:figure2}
\end{figure}

This can be seen by considering the quality factor of a resonator driven at frequency $\omega$,
defined as $Q = \omega (W_m+W_e)/P_l$, where $W_m$ and $W_e$ are the time-averaged
magnetic  and electric energies stored in a given volume, respectively, and $P_l$ is the
time-averaged power dissipated in that volume\cite{Pozar:2005}.  For a resonator driven on resonance,
$\omega=\omega_0$ and $W_m=W_e$, so that $Q = 2 \omega_0 W_e/P_l$. Furthermore, $P_l$ can be
expressed as $P_l = \frac{1}{2}\omega_0\left (\mathrm{Im}\int \epsilon |\vec{E}|^2 \, d^3x +
\mathrm{Im}\int \mu |\vec{H}|^2 \, d^3x \right )$, where $\epsilon$ is the spatially-varying
complex dielectric constant.  Ignoring magnetic loss, which we do not believe to contribute
significantly, this reduces to $P_l = \frac{1}{2}\omega_0 \mathrm{Im}\int \epsilon |\vec{E}|^2 \,
d^3x$.  With $W_e = \frac{1}{4} \textrm{Re} \int \epsilon |\vec{E}|^2 d^3x$, we can re-express the
resonant quality factor as
\begin{equation}
    Q = \frac{\mathrm{Re} \int \epsilon |\vec{E}|^2 \, d^3x}{\mathrm{Im} \int \epsilon |\vec{E}|^2 \, d^3x}.
\end{equation}
It is useful to consider the time-averaged electric energy divided by the quality factor,
\begin{equation}
\label{eq:ratio}
    \frac{W_e}{Q} = \frac{1}{4}\textrm{Im}\int \epsilon |\vec{E}|^2 d^3x.
\end{equation}
This is a general expression for a spatially-varying dielectric constant.  In our structures, the
total volume can be divided into distinct isotropic regions.

For example, for a CPW resonator formed by Al patterned on 100 nm SiN$_x$, on a
100 $\Omega$-cm single-crystal Si substrate, we separate Eq. (\ref{eq:ratio}) into two parts,
$W_e/Q_0 = \frac{1}{4}\mathrm{Im}\int_A \epsilon_A |\vec{E}_A|^2 \, d^3x +
\frac{1}{4}\mathrm{Im}\int_B \epsilon_B |\vec{E}_B|^2 \, d^3x$, where the volumes $A$ and $B$
correspond to the regions occupied by the SiN$_x$ and the Si, respectively. This can be re-written
as $W_e/Q_0 = W_{eA}/Q_A+W_{eB}/Q_B$, or in terms of the intrinsic loss tangents $\tan \delta_{i,A}$
and $\tan \delta_{i,B}$ as
\begin{equation}
\label{eq:SiN CPW}
    W_e \tan \delta_0 = W_{eA} \tan \delta_{i,A} + W_{eB} \tan \delta_{i,B}.
\end{equation}
A finite-element analysis of the electric field distribution shows 11$\%$ of the total
time-averaged energy is stored in the SiN$_x$, 81$\%$ in the Si, and the remainder in vacuum. An
independent measurement of the loss for SiN$_x$, using an $LC$ resonator, yielded $\tan
\delta_{i, SiN} = 1.8 \times 10^{-4}$.  The intrinsic loss tangent for single-crystal silicon was
extracted from the analysis of a CPW resonator on 100 $\Omega$-cm Si, yielding $\tan \delta_{i, Si}
= 4.8\times10^{-6}$. Using Eq. \ref{eq:SiN CPW}, we calculate an expected loss tangent for the
composite sample of $\tan\delta_{0,calc} = 2.3\times10^{-5}$. The measurement yielded
$\tan\delta_{0,exp} = 2.2\times10^{-5}$, in excellent agreement.

\begin{figure}
\includegraphics[width=0.45\textwidth]{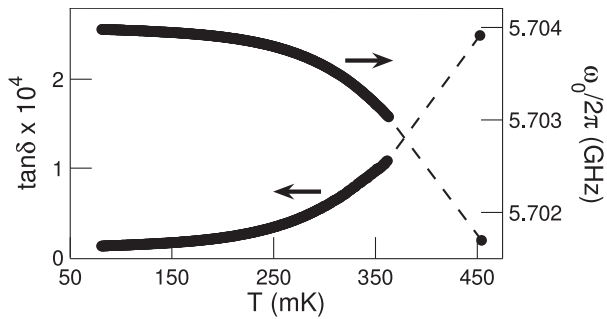} \caption{Temperature dependence of an
Al/100 $\Omega$-cm single-crystal Si CPW, measured with an rms excitation voltage of 1 mV.}
\label{fig:figure3}
\end{figure}

Equation \ref{eq:ratio} can also be used to extract the intrinsic loss of a given dielectric layer,
once the intrinsic loss tangents of the other layers in the structure are known. In this way we
extracted the intrinsic loss tangents of all the dielectrics measured with CPW resonators, as
tabulated in Table I. For example, a sample of Al/300nm SiO$_2$/100 $\Omega$-cm Si was used to
measure the loss of thermal SiO$_2$ that was commercially grown in a furnace.  The limiting loss
tangent of this CPW resonator was $8.9 \times 10^{-5}$; a finite-element analysis showed that
27$\%$ of energy is stored in the SiO$_2$, 61$\%$ in the Si, and 12$\%$ in the vacuum. We thus find
that the intrinsic loss tangent of thermal SiO$_2$ is $3.1 \times 10^{-4}$.

As expected, thermal SiO$_2$ exhibits comparatively high loss\cite{Liu:1997}. The results of Table
I imply that a more highly constrained lattice is correlated to lower loss.  This can be seen in
the silicon compounds where the transition from SiO$_2$ $\rightarrow $ SiN$_x$ $\rightarrow $ a-Si:H
$\rightarrow$ single-crystal Si corresponds to an increase in coordination number and a decrease in
loss.  Furthermore, the lower bounds on single-crystal Si and sapphire are not known precisely,
because the measurements may be limited by factors other than dielectric loss, such as radiation.
However, fabricating devices with single-crystal dielectrics is more difficult than using easily
deposited amorphous materials.  Due to this, we are currently optimizing the deposition of a-Si:H
since it is the least lossy amorphous material, and in general, the loss tangent has been seen to
correlate to the coherence times in our phase qubits\cite{Neeley:2008}.

These measurements were all taken at temperatures near 100 mK.  At higher temperatures the
dielectric loss may be overshadowed by the loss in the superconducting Al electrodes
\cite{Mattis:1958}. In Fig. 3 we display the temperature-dependent loss of an Al/100 $\Omega$-cm Si
CPW. The higher loss with increasing temperature, and the frequency shift, are consistent with
other measurements \cite{Mazin:2002,Mazin:2006}.

In conclusion, we have reported the low voltage, low temperature, intrinsic loss of many dielectrics.
Furthermore, we have shown how to extract the intrinsic dielectric loss from CPW resonator data and find the results
of measured CPW resonators to be commensurate with values given by LC resonators. Discovering other materials with lower
loss tangents than the dielectrics reported here would offer significant improvements in qubit
coherence times, and may be a crucial step in developing a scalable superconducting quantum
computer.

\textbf{Acknowledgements.} Devices were made at the UCSB and Cornell Nanofabrication Facilities, a part of the NSF-funded National Nanotechnology Infrastructure Network. This work was
supported by ARDA under grant W911NF-04-1-0204 and by the NSF under grant CCF-0507227.

\end{document}